\documentclass[useAMS,onecolumn,usenatbib]{mn2e}

\newcommand{\prt}{\partial}

\title[Linearized perturbation on stationary inflow solutions
in an inviscid and thin accretion disc]{Linearized perturbation on
stationary inflow solutions in an inviscid and thin accretion disc}
\author[Arnab K. Ray]{Arnab K. Ray\thanks{E-mail:
tpakr@mahendra.iacs.res.in}\\
Department of Theoretical Physics, Indian Association for the
Cultivation of Science, Jadavpur, Calcutta 700032, INDIA}
\begin{document}

%\date{}

%\pagerange{\pageref{}--\pageref{}} \pubyear{}

\maketitle

%\label{firstpage}

\begin{abstract}
The influence of a linearized perturbation on stationary inflow 
solutions in an inviscid and thin accretion disc has been studied here, 
and it has been argued, that a perturbative technique would indicate that
all possible classes of inflow solutions would be stable. The choice 
of the driving potential, Newtonian or pseudo-Newtonian, would not 
particularly affect the arguments which establish the stability
of solutions. It has then been surmised that in the matter of the 
selection of a particular solution, adoption of a non-perturbative
technique, based on a more physical criterion, as in the case of the
selection of the transonic solution in spherically symmetric accretion,
would give a more conclusive indication about the choice of a particular
branch of the flow. 
\end{abstract}

\begin{keywords}
accretion, accretion discs -- black hole physics -- hydrodynamics
-- methods: analytical
\end{keywords}

\section{Introduction}

The object of this work has been to study the stability of the
stationary inflow solutions in axially symmetric accretion, under 
the influence of a linearized perturbation. The chosen model has 
been that of an inviscid and thin accretion disc. 

Use of this model, in which viscous effects have been ignored, has
found a fairly widespread and consistent favour over the years
\citep{az81,chk89,msc96}. As a consequence of such a physical prescription,
there remains no mechanism for the outward transport of the angular
momentum of the flow, and hence the specific angular momentum of the
flow has been taken to be a constant of the motion. This naturally 
simplifies the analysis to a great extent, without significantly 
detracting from a true understanding of the underlying physics. The
whole analysis has been based on the thin-disc configuration in
accretion studies -- a very well-known textbook model \citep{pri81,fkr92}
that has been quite readily and regularly invoked by researchers in
accretion astrophysics. 

It has been argued here, that in an inviscid and thin accretion disc,
the stationary inflow solutions of abiding interest, are all stable 
under the influence of a linearized time-dependent perturbation. The
technique adopted has been similar to the one already existing for the
case of spherically symmetric accretion \citep{pso80,td92}. Separate
emphasis has been laid on discussing the stability of the so-called 
transonic inflow solution -- a solution that passes through the 
critical point(s) of the flow, and whose velocity is subsonic very
far away from the accretor, becoming supersonic as the accretor is
approached. The subsonic inflows, a class of physically meaningful
inflow solutions, the maximum of whose velocity always remains below
the velocity of the inflow solution passing through the critical
point(s), have been quite extensively considered as regards their
stability, since, like the transonic inflow, the
subsonic inflow solutions also obey the same outer boundary condition
that the flow velocity should decrease very far away from the accretor. 
And it may also be pointed out that since the subsonic solutions do not 
pass through any singular point of the flow, the mathematical problem 
of studying their stability is much simpler compared with that of 
the transonic solution. 

The classical Newtonian potential has been taken as the driving 
factor which effects the infall of matter, although it is easy to
see that the use of the pseudo-Newtonian potential of \citet{pw80}
-- which satisfactorily represents general relativistic
effects in a Newtonian framework and which is frequently in use to 
study the infall of matter under the gravitational influence of a
black hole -- alters nothing radically about the
conclusions drawn from a linear stability analysis of the inflow 
solutions. What it changes, however, is the number of critical 
points of the flow. In the context of the use of the Paczy\' nski-Wiita
potential, mention may be made here of the fact that the number of 
critical points of the flow is also determined by various constants of the 
motion like accretion rate, specific angular momentum and the specific
energy of the flow. It has been shown \citep{skc90} that for a given 
angular momentum of the flow, there exists a range of accretion rate,
or equivalently a range of energy, such that the flow has three critical
points. Of these, the two outer points are saddle points, flanking 
between themselves, a third centre-type point. The flow, however, has
only one critical point for values of accretion rate and energy which do
not lie within the given range that will develop three critical points. 

In solutions which admit of more than one saddle-type critical point,
the existence of shocks is a possibility \citep{skc90}. For accretion
on to a black hole, a solution would then pass through the outermost
critical point (which is a saddle point), undergo a shock and then
pass through the innermost critical point (again a saddle point) to 
cross the horizon with supersonic velocities. For accretion on to a 
neutron star, a flow passing through the inner critical point must 
undergo another shock to meet the inner boundary condition. The 
behaviour of these shocks are understood by invoking the Rankine-Hugoniot
conditions \citep{chk89}. Once again it is to be emphasized that the 
outer boundary condition for all kinds of physically meaningful inflow
solutions -- shocked or continuous -- is that the bulk velocity should 
decrease to zero very far away from the accretor. 

\section{The Equations of the Flow}

For the radial drift velocity $v$ and density $\rho$, the flow is 
governed by Euler's equation and the continuity equation. The radial
component of Euler's equation is given by

\begin{equation}
\label{e1}
{\frac{\prt v}{\prt t}}+v{\frac{\prt v}{\prt R}}
+{\frac{1}{\rho}}{\frac{\prt P}{\prt R}}+{\frac{\prt V}{\prt R}}
-{\frac{L^2}{R^3}}=0
\end{equation}
where $L$, which is a constant, is the specific angular momentum, and
the Newtonian gravity potential $V = -GM/R$. The choice of the
pseudo-Newtonian potential of \citet{pw80} would give $V = -GM/(R-{r_g})$, 
in which $r_g$ is the Schwarzschild radius of a black hole. 

Use has been made of a polytropic equation of state

\begin{equation}
\label{e2}
P=k{\rho}^{\gamma}
\end{equation}
where $\gamma$ is the polytropic exponent, whose admissible range    
$(1<{\gamma}<5/3)$ is restricted by the isothermal limit and the 
adiabatic limit respectively.

A surface density $\Sigma$ has been defined by vertically integrating
$\rho$ over the disc thickness $H$, and in the thin-disc approximation
\citep{fkr92} this gives 
${\Sigma}{\cong}{\rho}H$. The continuity equation, in 
terms of the surface density $\Sigma$, in the thin-disc approximation,
is thus given by

\begin{equation}
\label{e3}
{\frac{\prt \Sigma}{\prt t}}+{\frac{1}{R}}{\frac{\prt}{\prt R}}
\big({\Sigma}vR \big)=0
\end{equation}

Again invoking the thin-disc approximation, the disc thickness $H$
is obtained from the vertical component of Euler's equation as

\begin{equation}
\label{e4}
H{\cong}{\frac{c_s}{(GM)^{1/2}}}R^{3/2}
\end{equation}
where $c_s$, the speed of sound, obeys the relation 

\begin{equation}
\label{e5}
{c_s}^2={\gamma}k{\rho}^{\gamma -1}.
\end{equation}

The set of equations (\ref{e1})--(\ref{e5}), defines the whole problem 
completely. From (\ref{e1}) and (\ref{e2}) is obtained

\begin{equation}
\label{e6}
{\frac{\prt v}{\prt t}}+v{\frac{\prt v}{\prt R}}
+k{\gamma}{\rho}^{\gamma -2}{\frac{\prt \rho}{\prt R}}+
{\frac{\prt V}{\prt R}} -{\frac{L^2}{R^3}}=0
\end{equation}
while (\ref{e3}),(\ref{e4}) and (\ref{e5}) give

\begin{equation}
\label{e7}
{\frac{\prt}{\prt t}}\big[{\rho}^{(\gamma +1)/2}\big]+R^{-5/2}
{\frac{\prt}{\prt R}}\big[{\rho}^{(\gamma +1)/2}vR^{5/2}\big]=0
\end{equation} 

The two equations (\ref{e6}) and (\ref{e7}) above are those 
that govern the flow in the inviscid and thin accretion disc.

\section{The Perturbation Equation}

The steady-state solutions of (\ref{e6}) and (\ref{e7}) 
are $v_0$ and ${\rho}_0$,
on which are imposed small first-order perturbations $v^{\prime}$
and ${\rho}^{\prime}$ respectively. Following a similar prescription
by \citet{pso80} for spherically symmetric accretion, it is convenient
to introduce a new variable, $f{\equiv}{\rho}^{(\gamma +1)/2}vR^{5/2}$,
which in physical terms is the mass flux. Its steady-state value $f_0$,
as can be seen from (\ref{e7}), is a constant that is identified as the 
mass accretion rate. A linearized perturbation $f^{\prime}$, about $f_0$ 
is given by,

\begin{equation}
\label{e8}
f^{\prime}={{\rho}_0}^{(\gamma -1)/2}R^{5/2}\bigg(v^{\prime}{{\rho}_0}
+\frac{\gamma +1}{2}{v_0}{\rho}^{\prime}\bigg)
\end{equation}

Linearizing in the perturbation variables $v^{\prime}$ and ${\rho}^
{\prime}$, gives from (\ref{e7}), the expression,

\begin{equation}
\label{e9}
{\frac{\prt {{\rho}^{\prime}}}{\prt t}}+{\frac{2}{\gamma +1}}
{{\rho}_0}^{-(\gamma -1)/2}R^{-5/2}{\frac{\prt f^{\prime}}{\prt R}}=0
\end{equation}

With the use of (\ref{e9}), successive differentiation of (\ref{e8}) with 
respect to time yields, first

\begin{equation}
\label{e10}
{\frac{\prt v^{\prime}}{\prt t}}=R^{-5/2}{{\rho}_0}^{-(\gamma +1)/2}
\bigg({\frac{\prt f^{\prime}}{\prt t}}+{v_0}
{\frac{\prt f^{\prime}}{\prt R}}\bigg)
\end{equation}
and then

\begin{equation}
\label{e11}
{\frac{{\prt}^2 v^{\prime}}{\prt t^2}}=R^{-5/2}{{\rho}_0}^{-(\gamma +1)/2}
\bigg[{\frac{{\prt}^2 f^{\prime}}{\prt t^2}}+{v_0}{\frac{\prt}{\prt R}}
\bigg({\frac{\prt f^{\prime}}{\prt t}}\bigg)\bigg]
\end{equation}

Now linearizing in terms of the perturbation variables, gives 
from (\ref{e6}) the equation,

\begin{equation}
\label{e12}
{\frac{\prt v^{\prime}}{\prt t}}+{\frac{\prt}{\prt R}}\bigg({v_0}v^{\prime}
+{c_{s0}}^2{\frac{{\rho}^{\prime}}{{\rho}_0}}\bigg)=0
\end{equation}
where $c_{s0}$ is the speed of sound in the steady state. 

Differentiating (\ref{e12}) with respect to time, and using the conditions
given by (\ref{e9}),(\ref{e10}) and (\ref{e11}) will finally deliver the 
perturbation equation as 

\begin{equation}
\label{e13}
{\frac{{\prt}^2 f^{\prime}}{\prt t^2}}+2{\frac{\prt}{\prt R}}\bigg({v_0}
{\frac{\prt f^{\prime}}{\prt t}}\bigg)+{\frac{1}{v_0}}
{\frac{\prt}{\prt R}}\bigg[{v_0}\bigg({v_0}^2-{\frac{2}{\gamma +1}}
{c_{s0}}^2\bigg){\frac{\prt f^{\prime}}{\prt R}}\bigg]=0
\end{equation}

This expression bears a very close resemblance in form to a perturbation
equation obtained by \citet{td92}, using similar methods, for the case of
spherically symmetric accretion. 

\section{Stability Analysis of Solutions}

Before taking up the stability analysis of solutions, as implied 
by (\ref{e13}),
it would be worthwhile to have an understanding of the critical 
points of the flow. In the steady state, (\ref{e6}) and (\ref{e7}) 
can be combined
to give the critical points -- the required condition being that 
the numerator and the denominator of $dv_0/dR$ vanish simultaneously
at those points. This leads to the expressions

\begin{eqnarray}
\label{e14}
{v_0}^2 &=& \frac{2}{\gamma +1}{c_{s0}}^2  \nonumber \\
\frac{5}{\gamma +1}{c_{s0}}^2 &=& \frac{GM}{R}-\frac{L^2}{R^2}
\end{eqnarray}

The second equation above, being a quadratic in $R$, indicates that
there would be two critical points for the flow. The choice
of the pseudo-Newtonian potential of \citet{pw80}, merely changes 
the number of critical points \citep{chk89}, without altering the first 
of the two equations above. 

It is seen that the transonic inflow which originates very far away
from the accretor, with a very low subsonic velocity, will pass 
through the first point -- which is a saddle point, and
then tend to curl around the second point -- which is a centre-type 
point. In case the accretor is a large and distended object like
an ordinary star (which justifies the use of the conventional Newtonian
potential in this analysis), it is quite likely then that while the 
flow solution tends to curl around the inner critical point, it would
also meet the surface of the accretor. This, as will be discussed later, 
has important implications for the stability of the transonic inflow. 

Returning now to (\ref{e13}), a solution of the form 
$f^{\prime}=g(R)e^{- {\mathrm{i}} \Omega t}$ is chosen,
in which $g(R)$ is the spatial part of the perturbation and $\Omega$
is real. This leads to the result

\begin{equation}
\label{e15}
-g{\Omega}^2-2{\mathrm{i}}{\Omega}{\frac{d}{dR}}({v_0}g)
+{\frac{1}{v_0}}{\frac{d}{dR}}\bigg[{v_0}\big({v_0}^2-{\beta}^2
{c_{s0}}^2\big){\frac{dg}{dR}}\bigg]=0
\end{equation} 
where ${\beta}^2 = 2/(\gamma +1)$.

For subsonic flows, it is easy to understand that there are two
values of $R$ -- one close to the accretor and one very far away
-- where the perturbation, which is in the form of a standing wave,
can be constrained to die out. Multiplying (\ref{e15}) by ${v_0}g$ 
and integrating over the range of $R$, which is bounded by
the two points where the perturbation dies out, yields,

\begin{equation}
\label{e16}
{\Omega}^2
\int {v_0}g^2 \, dR = -\int {v_0}\big({v_0}^2-{\beta}^2{c_{s0}}^2\big)
\bigg( \frac{dg}{dR} \bigg)^2 \, dR
\end{equation}

The above result could be obtained because the boundary terms all
vanish. With a real value of $g$, for the subsonic flows -- all
of them governed everywhere by the 
condition ${v_0}^2<{\beta}^2{c_{s0}}^2$ 
-- ${\Omega}^2$ will be positive, implying that the perturbation, 
which is in the nature of a standing wave, will
be oscillatory and at least will not grow in time. This line of
reasoning was analogously established for spherically symmetric subsonic
flows by \citet{pso80}. 

It is also possible to subject the subsonic flow solutions to an analysis
where the perturbation may be treated as a travelling wave. The method 
for this, again follows a prescription used by \citet{pso80} for the
spherically symmetric flow. The perturbation would have to be
confined to a short-wavelength regime, in which the wavelength of the 
propagating waves will be much less compared to a characteristic 
length-scale of the system, which, in this case, as \citet{pso80} argued
for the analogous spherically symmetric situation, could
be the radius of the accretor, and that for an ordinary star is $R_{\star}$.
Since such a length-scale would be significantly large, it is somewhat
justifiable to restrict the study to a perturbation of comparatively
short wavelengths. Consequently, $\Omega$, the frequency of the waves,
would be large.

It is convenient to recast (\ref{e15}) as 

\begin{equation}
\label{e17}
\big({v_0}^2 - {\beta}^2{c_{s0}}^2\big)\frac{d^2 g}{dR^2}
+\bigg[3{v_0}{\frac{dv_0}{dR}} - {\beta}^2{\frac{1}{v_0}}
{\frac{d}{dR}}\big(v_0 {c_{s0}}^2\big) - 2{\mathrm{i}}
{\Omega}v_0 \bigg]\frac{dg}{dR}-\bigg(2{\mathrm{i}}{\Omega}
{\frac{dv_0}{dR}} + {\Omega}^2\bigg)g = 0
\end{equation}

A solution to (\ref{e17}) may be obtained by expanding $g(R)$ as a power
series \citep{pso80}, in the form 

\begin{equation}
\label{e18}
g(R)= \exp \Bigg[\sum_{n=-1}^{\infty} \frac{k_n(R)}{{\Omega}^n}\Bigg]
\end{equation}
and to solve for the first few terms. 

For such a solution, terms in $\Omega ^2$, $\Omega$ and $\Omega ^0$
will respectively yield the conditions

\begin{equation}
\label{e19}
\big({v_0}^2 - {\beta}^2 {c_{s0}}^2\big)\bigg({\frac{dk_{-1}}{dR}}\bigg)^2
-2{\mathrm{i}}{v_0}{\frac{dk_{-1}}{dR}} -1 = 0
\end{equation}

\begin{equation}
\label{e20}
\big({v_0}^2 - {\beta}^2 {c_{s0}}^2\big)\bigg({\frac{d^2 k_{-1}}{dR^2}}
+ 2{\frac{dk_{-1}}{dR}}{\frac{dk_0}{dR}}\bigg) 
+ \bigg[ 3{v_0}{\frac{dv_0}{dR}} - \frac{\beta ^2}{v_0}\frac{d}{dR}
\big({v_0}{c_{s0}}^2 \big) \bigg]\frac{dk_{-1}}{dR} 
- 2{\mathrm{i}}{v_0}\frac{dk_0}{dR} - 2{\mathrm{i}}\frac{dv_0}{dR} = 0
\end{equation}
and

\begin{equation}
\label{e21}
\big({v_0}^2 - {\beta}^2 {c_{s0}}^2\big)\Bigg[{\frac{d^2 k_0}{dR^2}}
+ 2{\frac{dk_{-1}}{dR}}{\frac{dk_1}{dR}} + \bigg({\frac{dk_0}{dR}}\bigg)^2
\Bigg]
+ \Bigg [3{v_0}{\frac{dv_0}{dR}}
- {\frac{\beta ^2}{v_0}}{\frac{d}{dR}}\big({v_0}{c_{s0}}^2\big)\Bigg]
\frac{dk_0}{dR} - 2{\mathrm{i}}{v_0}\frac{dk_1}{dR} = 0
\end{equation} 

It is to be noted that as compared to the spherically symmetric case, the 
analogous results for the inviscid and thin accretion disc, given by 
(\ref{e19}), (\ref{e20}) and (\ref{e21}), vary by a scale
factor of $\beta$ for the speed of sound in the steady state. It is 
then quite straightforward to arrive at the expressions,

\begin{equation}
\label{e22}
\frac{dk_{-1}}{dR} = \frac{{\mathrm{i}}}{v_0 \pm \beta c_{s0}}
\end{equation}
and

\begin{equation}
\label{e23}
k_0 = - {\frac{1}{2}} \ln \big(\beta v_0 c_{s0}\big) + {\mathrm{constant}}
\end{equation}

The first two terms in the power series expansion of $g(R)$ are given
by (\ref{e22}) and (\ref{e23}). As with the spherically symmetric case, 
here also self-consistency may be had by establishing 

\begin{equation}
\label{e24}
\Omega \vert k_{-1}(R) \vert \gg \vert k_0 \vert \gg {\Omega}^{-1}
\vert k_1 (R) \vert
\end{equation}

From the way it has been chosen, 
$\Omega \gg (v_0 \pm \beta c_{s0})/R_{\star}$, it can be seen that 
the condition given by (\ref{e24}) is satisfied, if the radial
distance is not too small. For large $R$, the asymptotic behaviour of 
$k_1$, $k_{-1}$ and $k_0$ obtained from (\ref{e21}), (\ref{e22}) and 
(\ref{e23}) respectively, is given by 

\begin{equation}
\label{e25}
k_1 \sim \frac{1}{R}\, , \quad k_{-1} \sim R \, , \quad k_0 \sim \ln R
\end{equation}

The result given above upholds the self-consistency required by the
condition in (\ref{e24}). The results in (\ref{e25}) follow from a very 
possible power-law dependence of the steady flow velocity on the radius, 
at great distances from the accretor.

The expression for the perturbation may then be written as 

\begin{equation}
\label{e26}
f^{\prime} = \frac{e^{-{\mathrm{i}} \Omega t}}{\sqrt{\beta v_0 c_{s0}}}
\bigg[A_{+} \exp \bigg({\mathrm{i}} \Omega \int 
\frac{dR}{v_0 + \beta c_{s0}} \bigg)
+ A_{-} \exp \bigg({\mathrm{i}} \Omega \int 
\frac{dR}{v_0 - \beta c_{s0}}\bigg)\bigg]
\end{equation}

In (\ref{e26}) there is a linear superposition of two solutions 
with arbitrary
constants $A_{+}$ and $A_{-}$, both travelling with velocity $\beta c_{s0}$,
relative to the fluid, which itself moves with a bulk velocity $v_0$. 
One solution moves along with the bulk flow, while the other moves 
against it. 

Use of (\ref{e9}) and (\ref{e26}) gives the perturbation in density as

\begin{equation}
\label{e27}
{\rho}^{\prime} = {\beta}^2 R^{-5/2} {{\rho}_0}^{-(\gamma -1)/2}
\frac{e^{-{\mathrm{i}} \Omega t}}{\sqrt{\beta v_0 c_{s0}}}
\bigg[{\frac{A_{+}}{v_0 + \beta c_{s0}}} 
\exp \bigg({\mathrm{i}} \Omega \int{\frac{dR}{v_0 + \beta c_{s0}}}\bigg)
+ {\frac{A_{-}}{v_0 - \beta c_{s0}}}
\exp \bigg({\mathrm{i}} \Omega \int{\frac{dR}{v_0 - \beta c_{s0}}}\bigg) 
\bigg]
\end{equation}

Using (\ref{e27}), the perturbation in velocity can then be obtained 
from (\ref{e8}), 
and following some simple algebraic manipulation it can be rendered as  

\begin{equation}
\label{e28}
v^{\prime} = \pm \frac{c_{s0}}{\beta} \frac{{\rho}^{\prime}}{{\rho}_0}
\end{equation}
in which the positive and negative signs indicate outgoing and 
incoming waves, respectively. 

In a unit volume of the fluid, the kinetic energy content is

\begin{equation}
\label{e29}
{\mathcal{E}}_{\mathrm{kin}} = \frac{1}{2}\big({\rho}_0 
+ {\rho}^{\prime}\big)\big(v_0 + v^{\prime}\big)^2
\end{equation}

The potential energy per unit volume of the fluid is the sum of the 
gravitational energy, the rotational energy and the internal energy, 
and is given by

\begin{equation}
\label{e30}
{\mathcal{E}}_{\mathrm{pot}} = \big({\rho}_0 + {\rho}^{\prime}\big)
{\frac{GM}{R}}
- \big({\rho}_0 + {\rho}^{\prime}\big){\frac{L^2}{2R^2}} 
+ {\rho}_0 \epsilon + {\rho}^{\prime}{\frac{\prt}{\prt {\rho}_0}}
\big({\rho}_0 \epsilon\big) + \frac{1}{2}{{\rho}^{\prime}}^2 
{\frac{{\prt}^2}{\prt {{\rho}_0}^2}}\big({\rho}_0 \epsilon \big)
\end{equation}
where $\epsilon$ is the internal energy per unit mass \citep{ll87}.

The first-order terms vanish on time averaging. In that case, the 
contribution to the total energy in the perturbation comes from the
second-order terms, which is given by

\begin{equation}
\label{e31}
{\mathcal{E}}_{\mathrm{pert}} = \frac{1}{2}{\rho}_0{v^{\prime}}^2 
+ v_0 {\rho}^{\prime}v^{\prime} +  \frac{1}{2}{{\rho}^{\prime}}^2
{\frac{{\prt}^2}{\prt {{\rho}_0}^2}}\big({\rho}_0 \epsilon \big)
\end{equation}

For an adiabatic perturbation, such as the one being prescribed here, 
the condition $ds=0$ can be imposed on the thermodynamic 
relation, $d{\epsilon}= Tds + \big(P/{\rho}^2 \big)d{\rho}$, which 
will then give 

\begin{equation}
\label{e32}
{\frac{{\prt}^2}{\prt {{\rho}_0}^2}}\big({\rho}_0 \epsilon \big)
{\bigg{\vert}}_s = \frac{{c_{s0}}^2}{{\rho}_0}
\end{equation}

Combining the results from (\ref{e28}),(\ref{e31}) and (\ref{e32}) 
will give

\begin{equation}
\label{e33}
{\mathcal{E}}_{\mathrm{pert}} = 
\frac{c_{s0}}{{\beta}^2}\frac{{{\rho}^{\prime}}^2}{{\rho}_0}
\bigg[\frac{c_{s0}}{2}\big({\beta}^2 + 1 \big) \pm \beta v_0 \bigg]
\end{equation}

The expression for ${\rho}^{\prime}$ is to be obtained from (\ref{e27}), 
and upon time averaging, the result will be

\begin{equation}
\label{e34}
{\mathcal{E}}_{\mathrm{pert}} = \frac{1}{2}\frac{\beta A^2}{f_0} 
\frac{R^{-5/2} {{\rho}_0}^{-(\gamma -1)/2}}{\big(v_0 \pm \beta 
c_{s0}\big)^2}
\bigg[\frac{c_{s0}}{2}\big({\beta}^2 + 1 \big) \pm \beta v_0 \bigg]
\end{equation}
where $f_0 \equiv {{\rho}_0}^{(\gamma +1)/2} v_0 R^{5/2}$, is a constant
that in physical terms is the steady-state value of the accretion rate.
The factor $1/2$ in (\ref{e34} ) derives from 
the averaging of ${{\rho}^{\prime}}^2$.

The total energy flux in the perturbation is obtained by multiplying 
${\mathcal{E}}_{\mathrm{pert}}$
by the propagation velocity $(v_0 \pm \beta c_{s0})$ and then by 
integrating over the area of the cylindrical face of the accretion
disc, which is $2 \pi RH$. Here $H$ is to be substituted from (\ref{e4}) 
and (\ref{e5}), and together with the fact that $H \ll R$ in the 
thin-disc approximation \citep{fkr92}, an expression for the 
energy flux will be delivered as

\begin{equation}
\label{e35}
{\mathcal{F}} = \frac{\pi A^2}{f_0}\bigg(\frac{\gamma k}{GM}\bigg)^{1/2} 
{\beta}^2 \bigg[\pm 1 + \frac{1- {\beta}^2}{2 \beta 
({\mathcal{M}} \pm \beta )}\bigg]
\end{equation}
where ${\mathcal{M}} = {v_0}/{c_{s0}}$, is the Mach number.

It is to be seen that for purely subsonic 
flows, ${\mathcal{M}}$ always
remains less than $\beta$, which is a condition that is easily 
understood from the first of the two critical point conditions, given
by (\ref{e14}). This would then imply that the total energy in the wave 
remains finite as the wave propagates. Indeed, for very great radial
distances, when ${\mathcal{M}} \longrightarrow 0$, 
the flux is less compared to what it is near the outer 
critical point of the flow. The perturbation does not manifest itself
as a runaway increase in the energy of the system. The system would 
then remain stable under the influence of such a perturbation.

For the case of the transonic flow, a different line of reasoning would
have to be adopted, since the condition ${v_0}^2<{\beta}^2{c_{s0}}^2$
would not hold everywhere, with the flow passing through a singular 
point. The physically meaningful flow in this case
will be subsonic beyond the saddle-type outer critical point, pass
through it to attain a supersonic velocity and then tend to curl around
the inner centre-type critical point, but not actually flow through
it. \citet{td92} have established the stability of the transonic solution
in spherical symmetry by arguing that the stability of the subsonic 
region of the flow, with the critical point as one boundary, will
depend on conditions at the far outer end of the flow. In so far as the
thin accretion disc is dependent only on $R$ (the disc being inviscid
and hence, the related equations are independent of any angle variable),
the analogous argument of \citet{td92} is expected to hold good to
ensure the stability of the subsonic region of the transonic solution.

For the supersonic region, which is finite, the reasoning of \citet{gar79}
that a disturbance in this region will be carried away in a finite time,
will also ensure the stability of the flow in this region.              

There is another important argument to uphold the stability of the
transonic flow in a thin accretion disc. In the case of the spherically
symmetric transonic flow, the velocity near the accretor has an unbridled
growth, such that any perturbation on it cannot be constrained to die out
by a conceivable physical mechanism \citep{pso80}. In the inviscid and 
thin accretion disc, the transonic solution has no such runaway growth in
the inner region of the flow, tending to curl around the inner critical 
point as it does. And while doing so, it may reach the surface of the 
accretor, if the accretor is as distended as an ordinary star. This can 
ensure that a 
perturbation on the solution in this region will not behave in an
unconstrained manner.

A note of caution has to be sounded here. The discussion presented above, 
pertains only to flows which are continuous everywhere. However, for
spherically symmetric flows with discontinuities like shocks, \citet{td92}
have argued that the expectation would also be one in favour of the stability
of those flows. Under the present assumptions governing an inviscid and 
thin accretion disc, the mathematical problem becomes very similar to the 
case of spherically symmetric accretion. In that case the argument of 
\citet{td92} for spherical symmetry could very likely be extended to the
axially symmetric system being considered here. In any case, regardless
of the geometry of the problem, it is a matter of general understanding
that a disturbance in the supersonic region in a shocked solution,
is transmitted completely through the shock and that in the subsonic
region a disturbance is reflected at the shock, with a diminished 
reflected amplitude for the disturbance as compared with the incident 
amplitude \citep{td92}. More to the point, by a local stability analysis,
it has been demonstrated \citep{chk89} that Rankine-Hugoniot shocks in an
inviscid and thin accretion disc are stable under the influence of a
short-wavelength perturbation in real time. 

It can then be safely established that a physically well-behaved and 
meaningful inflow solution is indeed stable under the influence of  
a linearized perturbation. This conclusion holds good for both the
transonic inflow solution, as well as the entire class of subsonic
inflow solutions. 

\section{Concluding Remarks}

So far, it has been demonstrated that linear stability analysis,
which is essentially a perturbative technique, offers no clue about
the exclusive choice of a particular solution -- indeed, to the extent
that a perturbative technique can be relied upon, every solution
(transonic or subsonic) seems just as much realizable as any other.
It is then to be conjectured that the criterion for the selection of 
a particular solution would have to be non-perturbative in character
and would have to be based on fundamentally physical arguments, like
the maximization of the accretion rate or the minimization of the 
total energy associated with a solution, as in the case of the transonic
inflow solution in spherically symmetric accretion \citep{bon52,gar79}. 

A recent work \citep{rb02} on spherically symmetric 
accretion has indicated that such an approach, however,
would necessitate the addressing of the whole issue of the selection of
a particular solution, from the viewpoint of the temporal evolution of
the flow, instead of considering it solely on the basis of the 
stationary picture. The study has shown that it is the transonic 
solution which is decisively preferred to all the others. Such an
expectation would also be valid for the case of the inviscid and thin
accretion disc being discussed here.

Support for such a contention can also be had from a very different
quarter. Study carried out over the last two decades has established
that there is a very close one-to-one correspondence between certain
features of black hole physics and the physics of supersonic acoustic
flows. More specifically, for an inviscid, barotropic and irrotational
fluid flow (such as the conventional spherically symmetric accretion) 
the equation
describing an acoustic disturbance can be rendered in the form of a 
metric that relates very closely to the Schwarzschild metric, represented
in the Painlev\'er-Gullstrand form \citep{vis97}. Close to the sonic
point of the flow, the similarity is particularly evident. Such a 
close correspondence between the unidirectionality of trajectories
near a black hole and of flows passing through the sonic point, is a
reason strong enough to suggest that transonic trajectories are the 
favoured ones. Extending that argument, a comparison may also be 
drawn for rotational, inviscid fluid flows, to have an understanding 
of the favoured status of the trajectories in this case.

\section*{Acknowledgments}

This research has made use of NASA's Astrophysics Data System.
The financial assistance provided by the Council of Scientific and
Industrial Research, Government of India, is being gratefully 
acknowledged. Gratitude is also to be expressed to Prof. J. K.
Bhattacharjee for his very helpful advice, given unreservedly 
throughout this work.

\bsp

\label{lastpage}

\end{document}